\documentclass[12pt]{article}
\usepackage{graphicx,verbatim,array,multicol, subfigure, color, lscape, mathrsfs}
\usepackage{psfrag, amsmath, amsfonts, epsfig, fancybox, biometri,setspace,soul, cancel, amsthm}

\def\bZ{{\bf Z}}

\def\transpose{{\sf \scriptscriptstyle{T}}}

\def\trans{^{\transpose}}

\definecolor{brown}{RGB}{150,100,100}
\definecolor{green}{RGB}{000,150,100}
\definecolor{purple}{RGB}{150,000,180}

\def\blue{\color{blue}}
\def\red{\color{red}}

\def\red{\color{red}}

\def\blue{\color{blue}}

\def\Ssc{\mathcal{S}}

\def\bq{\mathbf{q}}

\def\bQ{\mathbf{Q}}

\def\blue{\color{blue}}
\def\tcbox#1{\vskip1mm \begin{center}
        \hspace{.0\textwidth}\vbox{\hrule\hbox{\vrule\kern6pt
\parbox{.95\textwidth}{\kern6pt \blue #1 (TC)\vskip6pt}\kern6pt\vrule}\hrule}
        \end{center} \vskip-5mm}

\def\myboxlayla#1{\vskip1mm \begin{center}
        \hspace{.0\textwidth}\vbox{\hrule\hbox{\vrule\kern6pt
\parbox{.95\textwidth}{\kern6pt \red #1\vskip6pt}\kern6pt\vrule}\hrule}
        \end{center} \vskip-5mm}

\def\supA{^{(A)}}
\def\supB{^{(B)}}
\def\supg{^{(g)}}

\def\Isc{\mathcal{I}}

\def\Fhat{\widehat{F}}
\begin{document}

\title{Evaluating Surrogate Marker Information using Censored Data}

\author{Layla Parast$^{1}$, Tianxi Cai$^{2}$, Lu Tian$^{3}$ \\
$^{1}$RAND Corporation, 1776 Main Street, Santa Monica, CA 90401, parast@rand.org\\
$^{2}$Harvard University,  Department of Biostatistics, Boston, MA 02115 \\
$^{3}$ Stanford University, Department of Health, Research and Policy, Stanford, CA 94305
}

\date{}

\maketitle
\clearpage
\begin{abstract}
Given the long follow-up periods that are often required for treatment or intervention studies, the potential to use surrogate markers to decrease the required follow-up time is a very attractive goal. However, previous studies have shown that using inadequate markers or making inappropriate assumptions about the relationship between the primary outcome and surrogate marker can lead to inaccurate conclusions regarding the treatment effect. Currently available methods for identifying and validating surrogate markers tend to rely on restrictive model assumptions and/or focus on uncensored outcomes. The ability to use such methods in practice when the primary outcome of interest is a time-to-event outcome is difficult due to censoring and missing surrogate information among those who experience the primary outcome before surrogate marker measurement. In this paper, we propose a novel definition of the proportion of treatment effect explained by surrogate information collected up to a specified time in the setting of a time-to-event primary outcome. Our proposed approach accommodates a setting where individuals may experience the primary outcome before the surrogate marker is measured. We propose a robust nonparametric procedure to estimate the defined quantity using censored data and use a perturbation-resampling procedure for variance estimation. Simulation studies demonstrate that the proposed procedures perform well in finite samples. We illustrate the proposed procedures by investigating two potential surrogate markers for diabetes using data from the Diabetes Prevention Program.
\vspace{.1in}

\end{abstract}
\clearpage

\section{Introduction}
A surrogate marker is often defined as a physical measurement such as a biomarker, clinical measurement, or psychological test that can be ``used in therapeutic trials as a substitute for a clinically meaningful endpoint that is a direct measure of how a patient feels, functions, or survives and is expected to predict the effect of the therapy."(\cite*{temple1999surrogate}) The quest to identify, validate, and use surrogate markers in practice is driven by the potential for such markers to reduce the length of studies that currently require very long follow-up periods. For example, studies with outcomes such as cancer diagnosis, diabetes diagnosis, or heart attack often require many years of follow-up to precisely estimate an intervention effect. Research involving the identification and validation of surrogate markers has been active with much novel methodological development and heated debate. In a landmark paper, \cite{prentice1989surrogate} introduced a criterion for a valid surrogate marker which required that a test for treatment effect on the surrogate marker is also a valid test for treatment effect on the primary outcome of interest. Since then, numerous approaches have been developed to identify and validate surrogate markers and to quantify the ``surrogacy" of such markers. For example, motivated by the Prentice criterion, \cite{freedman1992statistical} proposed to estimate the proportion of treatment effect that is explained by a surrogate marker by examining the change in the regression coefficient for treatment when the surrogate marker is added to a specified regression model. \cite{wang2002measure} proposed a  more flexible model-based approach to estimate the proportion of treatment effect explained by defining a quantity that attempts to capture what the the effect of the treatment in the treatment group would be if the values of the surrogate were distributed as those in the control group.  
Building from the definition of \cite{wang2002measure}, \cite{parast2016} proposed a robust estimation procedure to estimate this quantity without the requirement of correct model specification.  However, none of these approaches are able to adequately accommodate settings with time-to-event outcomes.

While the proportion of treatment effect explained by a surrogate marker is intuitively appealing, a number of other quantities to assess surrogate markers have been proposed. For example, relative effect and adjusted association (\cite*{buyse1998criteria}), indirect and direct effects (\cite*{robins1992identifiability}), dissociative effects, associative effects, average causal necessity, average causal sufficiency, and the causal effect predictiveness surface in a principal stratification framework (\cite*{frangakis2002principal,huang2011comparing,gilbert2008evaluating,conlon2014surrogacy})  are some of the alternative quantities that are available.  Again however, the majority of these currently available methods were developed for settings where the primary outcome is fully observable and cannot be easily extended to settings with time-to-event outcomes.

For a survival outcome, $T$, existing methods largely require restrictive model assumptions that may not hold in practice.  \cite{lin1997estimating} extended the approach proposed by \cite{freedman1992statistical} to the survival setting but showed that it is actually impossible for both of the two specified survival models to hold simultaneously. The survival setting is often further complicated by the fact that the surrogate marker, $S$, itself may be missing. That is, the individual may be censored or may experience the primary outcome before $S$ is measured, in which case $S$ will not be observable and thus, commonly used metrics for surrogacy would not be well-defined.  When both $T$ and $S$ are subject to censoring, \cite{ghosh2008semiparametric} proposed estimation and inference procedures for the proportion of treatment effect explained by a surrogate using an accelerated failure time (AFT) model  and demonstrated desirable finite sample performance  when the AFT model holds.  \cite{ghosh2009assessing} proposed estimates of quantities to assess the validity of a surrogate marker in a semi-competing risks framework such that estimates are derived based on specified copula model and AFT models. However, their simulations demonstrate that when the assumed copula model is misspecified, the proposed procedure leads to biased estimates that persist with large sample sizes. In the principal stratification framework, multiple quantities for evaluating potential surrogate markers in a time-to-event outcome setting have been proposed by \cite{conlon2014surrogacy2}, using a Gaussian copula model with a Bayesian estimation approach. \cite{gabriel2014evaluating} and \cite{gabriel2015comparing} rely on flexible yet still model-based procedures using Weibull time-to-event models for the primary outcome. These methods would yield estimates that are difficult to interpret under model mis-specifications. 
%
It is thus of great interest to investigate methods that do not rely on correct model specification and are applicable to the survival setting where both $T$ and $S$ are subject to censoring. 

In this paper, we generalize the work of  \cite{parast2016} and propose a novel {model-free} framework for quantifying the proportion of treatment effect explained by surrogate information collected up to a specified time, $t_0$, in the survival setting. In addition, we propose a robust nonparametric procedure to estimate the defined quantity using censored data and a perturbation-resampling procedure for inference.  To increase efficiency, we also propose parallel augmented estimates that take advantage of baseline covariate information. Our proposed definition and estimation procedure is the only available method to accommodate survival settings where individuals may experience the primary outcome of interest or be censored before the surrogate marker is measured, a situation that is quite common in practice. We perform a simulation study to examine the finite sample performance of our proposed procedures and  illustrate the proposed procedures by investigating two potential surrogate markers for diabetes using data from the Diabetes Prevention Program.

\section{Setup and Definitions in a Causal Inference Framework \label{causal}}
Let $G$ be the binary treatment indicator with $G=A$ for treatment A and $G=B$ for treatment B and we assume throughout that subjects are randomly assigned to a treatment group at baseline. Let $T$ denote the survival time of interest and $S$ denote the surrogate marker value measured at time $t_0$.  Without loss of generality, we assume that $S$ only takes positive values (if not, we may simply exponentiate $S$). To study this problem under the causal inference framework, we use potential outcomes notation such that $T^{(g)}$ and $S^{(g)}$ denote the survival time and the surrogate marker value under treatment $G = g$. That is, $T^{(A)}, T^{(B)}, S^{(A)}$ and $S^{(B)}$ denote the survival time under treatment A, survival time under treatment B, surrogate marker value under treatment A and surrogate marker value under treatment B, respectively. We assume  that $S\supB$ and $S\supA$ have the same support. In practice, we can only { potentially} observe $(T, S)=(T^{(A)}, S^{(A)})$ or $(T^{(B)}, S^{(B)})$ {for each individual} depending on whether $G=A$ or $B.$ Throughout, we define the treatment effect, $\Delta(t)$, as the difference in survival { rates by time $t$} under treatment A versus under treatment B,
  $$ \Delta(t)=E\{ I(T^{(A)}>t)\} - E\{I(T^{(B)}>t)\} = P(T^{(A)}>t) - P(T^{(B)}>t)$$
where $t>t_0$. We consider a setting where individuals may be censored or experience the primary outcome before $t_0$ and thus, $S$ may not be observable for these patients. For simplicity, we assume that the surrogate marker cannot be measured after the primary outcome occurs, which is a reasonable assumption if, for example, the primary outcome is death, but discuss this assumption further in the Remarks.

Our analytic objective is to study the extent to which the surrogate information available at time $t_0$ captures the true treatment effect $\Delta(t)$. It is important to consider whether information concerning the primary outcome observed before $t_0$ should be considered as part of the surrogate information available at $t_0$. We argue that this information should indeed be considered as part of the surrogate information. That is, in this paper, we define surrogate information at $t_0$ as the combination of primary outcome information before $t_0$ and surrogate marker measurements collected at $t_0$ for those still being observed. We take this approach because even in the highly optimistic situation where one were to identify $S$, measured at $t_0$, as a valid surrogate marker that can be used to estimate and test for a treatment effect, it is unlikely that one would completely disregard primary outcome information that is observed up to $t_0$.  It is more sensible to envision that one uses \textit{both} primary outcome information before $t_0$ and surrogate marker measurements at $t_0$ to estimate the treatment effect, thus this combination of information is our definition of surrogate information at $t_0$ throughout this paper. Specifically, we consider the surrogate information available at $t_0$ as
$$
\bQ_{t_0} \supg = \{ T \supg \wedge  t_0, S\supg I(T \supg > t_0)\}, g=A, B.
$$
Further motivating this definition is the fact that it is difficult to consider a reasonable alternative to this approach. For example, one potential alternative would be to restrict estimation to only those who are still under observation at $t_0$ (i.e. removing individuals who experience the primary outcome before $t_0$) (\cite*{lin1997estimating,gabriel2015comparing}). However, it does not seem desirable to assess surrogacy on only a selected subset of survivors systematically different from the original population, nor does it seem reasonable to disregard observed information on $T$ before $t_0$ when the goal is to quantify the treatment effect on $T$.

We aim to define the proportion of treatment effect explained by $\bQ_{t_0}$ by contrasts between the actual treatment effect and the residual treatment effect that would be observed if the surrogate information available at $t_0$ under treatment A was equal to the surrogate information available at $t_0$ under treatment B. That is, we define the residual treatment
effect  by noting that
\begin{eqnarray*}
&&E\left\{ I(T \supA > t) - I(T \supB>t)  \mid \bQ_{t_0}\supA = \bQ_{t_0}\supB = \bq_{t_0}\right\}
= I(s_{t_0} >0)  \Delta_S^+(t, t_0, s_{t_0})  
\end{eqnarray*}
where $\bq_{t_0} = \{u_{t_0},  s_{t_0} \}=\{u\wedge t_0, sI(u>t_0)\}$,   and
$$
\Delta_S^+(t,t_0,s) = E\{  I(T \supA > t) - I( T\supB>t) \mid S\supA =S\supB=s, T\supA> t_0, T\supB > t_0\}.
$$
Thus, $\Delta_S(t,t_0,s_{t_0}) { = I(s_{t_0} >0)  \Delta_S^+(t, t_0, s_{t_0})}$ defines the hypothetical difference in survival at $t$  if the surrogate information available at time $t_0,$ $\bQ_{t_0}$, in both treatment groups was identical to $\bq_{t_0}$. Interestingly, $\Delta_S(t,t_0,s_{t_0})$ only depends on $\bq_{t_0}$ through $s_{t_0}$. An equivalent interpretation of $\Delta_S(t,t_0,s_{t_0})$ would be the hypothetical difference in survival at $t$ if \textit{both} the survival distribution up to $t_0$ and the distribution of the surrogate marker at
$t_0$ { among those who survived to $t_0$} were the same in the two treatment groups. This quantity summarizes the residual treatment effect that cannot be explained by the surrogate information available at $t_0$ {  and would be expected to equal zero for a perfect surrogate marker. }

However, $\Delta_S^+(t,t_0, s)$ is generally not identifiable since $S^{(A)}$ and $S^{(B)}$ cannot both be observed simultaneously.
To overcome this difficulty, we further assume that
\begin{eqnarray}
&&T^{(A)} \perp S^{(B)}(T\supB>t_0) \mid   \{ S^{(A)}, T\supA> t_0 \}\label{assumption1}\\
&&T^{(B)} \perp S^{(A)}(T\supA>t_0) \mid   \{ S^{(B)}, T\supB> t_0 \}\label{assumption2}
\end{eqnarray}
 Under assumptions (\ref{assumption1}) and (\ref{assumption2}),
\begin{align*}
\Delta_S^+(t,t_0, s_{t_0})
                  = \psi_A(t\mid s_{t_0}, t_0)-\psi_B(t \mid s_{t_0}, t_0)
\end{align*}
where for $g=A, B,$
$$\psi_g(t\mid s,t_0) = P(T^{(g)}> t\mid S^{(g)}=s, T\supg> t_0).$$
\def\Usc{\mathcal{U}}
\def\Ssc{\mathcal{S}}

To consider the residual treatment effect in a population, we may consider  $s_{t_0}$ as a realization from a random distribution $\Ssc_{t_0}$ and define the residual treatment effect as
$$
{  \Delta_S(t,t_0) = E\{\Delta_S(t,t_0,\Ssc_{t_0})\} = P(\Ssc_{t_0} > 0)E \left\{\Delta_S^+(t,t_0,\Ssc_{t_0}) \right\}.}
$$
The choice of the distribution of { $\Ssc_{t_0}$} depends on the specific context. For example, if treatment B is a placebo, then we may be interested in examining the residual treatment effect quantity when treatment A has no effect on the surrogate marker information at $t_0$, i.e., when the distribution of the surrogate information at $t_0$ under treatment $A$ is the same as that under treatment $B$. In this case,
$${ \Ssc_{t_0}} \sim I(T\supB > {t_0}) S\supB$$ and
\begin{align}
\Delta_S(t,t_0) = & P(T\supB >  t_0)  E\left\{  \Delta_S^+(t,t_0,S^{(B)}) \mid T_B > t_0\right\} \\ 
	     = &  P(T\supB > t_0)\left[ E\left\{  \psi_A(t \mid S^{(B)}, t_0)  \mid T_B > t_0\right\}  - E\left\{  \psi_B(t \mid S^{(B)}, t_0)  \mid T_B > t_0\right\}   \right]  \nonumber \\
              = & P(T\supB > t_0)\left\{\int \psi_A(t \mid s, t_0) dF_B(s \mid t_0)-P(T\supB> t\mid T\supB> t_0)\right\},
\label{surrogacyone}
\end{align}
where $F_g(\cdot\mid t_0)$ is the cumulative distribution function of $S^{(g)}$ conditional on $T\supg> t_0.$ Here treatment group $B$, the placebo group, serves as the reference distribution for the definition of $\Delta_S(t,t_0)$. Alternatively, when neither treatment A nor treatment B is a natural reference group, one may be interested in examining the residual treatment effect when the distribution of the surrogate information at $t_0$ in both groups are identical to that of a mixture population from the two groups. For example, in this case we may let
$${ \Ssc_{t_0}} \sim \frac{1}{2} I(T\supA > {t_0}) S\supA +\frac{1}{2} I(T\supB > {t_0})  S\supB$$
and
\begin{equation}
\Delta_S(t,t_0) = \int \Delta_S^+(t,t_0,s) d\left\{\frac{1}{2}P(T\supA > t_0)F_A(s \mid t_0)+\frac{1}{2} P(T\supB > t_0)F_B(s \mid t_0)\right\}. \label{surrogacytwo}
\end{equation}

{ For a given choice of $\Delta_S(t, t_0)$,} the proportion of treatment effect explained by the surrogate marker can be expressed
using a contrast between $\Delta(t)$ and $\Delta_S(t,t_0)$:
 \begin{equation} R_S(t,t_0)=\{\Delta(t)-\Delta_S(t,t_0)\}/\Delta(t)=1-\Delta_S(t,t_0)/\Delta(t). \label{defsm}\end{equation}
 In this paper, we focus on nonparametrically estimating this proportion using censored data.  Informally, we use $R_S(t,t_0)$ to measure the extent to which the surrogate marker captures information about the treatment effect on survival by comparing the total treatment effect with the hypothetical treatment effect when there is no difference in surrogate information up at $t_0$. The approach to defining the proportion of treatment effect explained based on contrasts between the actual treatment effect and the residual treatment effect was proposed in a non-survival setting in  \cite{wang2002measure} and further developed in  \cite{parast2016}. Our definition generalizes their proposed approach to a setting where the outcome may be a time-to-event outcome, individuals may be censored, and individuals may not have surrogate marker information available because they either experienced the primary outcome or were censored before the time of surrogate marker measurement.

\textit{Remark.}  This definition of the proportion of the treatment effect explained by the surrogate marker does not guarrantee that the resulting $R_S(t, t_0)$ is always between 0 and 1.  However, one set of sufficient conditions similar to those given in \cite{wang2002measure} is
\begin{enumerate}
\item[] (C1) $\psi_A(t|s, t_0)$ is a monotone increasing function of $s;$
\item[] (C2) $P(S\supA >s, T\supA>t_0)\ge P(S\supB >s, T\supB>t_0)$ for all $s;$
\item[] (C3) $\psi_A(t|s, t_0)\ge \psi_B(t|s, t_0) $ for all $s;$
\end{enumerate}
where the first condition implies that the surrogate marker at time $t_0$ is ``positively" related to the survival time; the second condition implies that there is a positive treatment effect on the surrogate marker and the third condition suggests that there is a non-negative residual treatment effect beyond that on the surrogate marker. { For (C1), $1/S$ can be used to replace $S$ if the surrogate markers is ``negatively" associated with the survival time.} 
In Appendix A in the Supplementary Materials, we show that under conditions (C1)-(C3), $0\le \Delta_S(t,t_0) \le \Delta(t)$ and $0\le R_S(t, t_0)\le 1$.

\section{Nonparametric Estimation of the proportion of treatment effect explained \label{estimation}}
The quantity $R_S(t,t_0)$ depends on the selection of the reference distribution $\Ssc_{t_0}$.  When the treatment group $B$ represents standard care or a placebo, the $\Delta_S(t,t_0)$ definition (\ref{surrogacyone}) seems intuitive. However, when neither group is a natural reference group, a new distribution may be considered such as the one used for the $\Delta_S(t,t_0)$ definition (\ref{surrogacytwo}). For simplicity,  we focus on the development of an estimation and inference procedure for the definition based on (\ref{surrogacyone}); however, parallel procedures would be applicable for other choices of a reference distribution.

Due to censoring, the observed data consist of $\{(X_{gi}, \delta_{gi}, S_{gi}), i=1,...,n_g; g = A, B\}$, where $X_{gi} = \min(T_{gi}, C_{ gi})$,  $\delta_{gi} = I(T_{gi} < C_{gi})$, $C_{gi}$ denotes the censoring time, and $S_{gi}$ denotes the surrogate marker information measured at time $t_0$, for $g= A,B$, for individual $i$. We assume that $(T_{gi}, S_{gi}) \perp
C_{gi}$. Throughout, we estimate the treatment effect $\Delta(t) =P(T^{(A)}>t) - P(T^{(B)}>t)$ as
$$ \widehat{\Delta}(t)  = n_A^{-1} \sum_{i=1}^{n_A} \frac{I(X_{Ai}>t)}{\widehat{W}^C_A(t)} - n_B^{-1} \sum_{i=1}^{n_B} \frac{I(X_{Bi}>t)}{\widehat{W}^C_B(t)}$$
where  $\widehat{W}^C_g(\cdot)$ is the Kaplan-Meier estimator of survival for censoring for $g=A,B.$ Note that this estimator is asymptotically equivalent to the difference of two Kaplan-Meier estimators for the survival time (see Appendix B in the Supplementary Materials).

\subsection{Nonparametric estimator of the proportion of treatment effect explained}

To estimate $\Delta_S(t,t_0)$ as defined in (\ref{surrogacyone}), we need to estimate
$${
\psi_A(t\mid s, t_0) = P(T^{(A)}>t\mid T^{(A)}>t_0, S^{(A)}=s)  \ \mbox{and} \ F_B(s \mid t_0) = P(S^{(B)} \le s \mid T^{(B)} > t_0) .}
$$
Note that $\psi_A(t\mid s,t_0) = P(T_{Ai}>t\mid X_{Ai}>t_0, S_{Ai}=s)$
given our earlier assumption that $(T_{gi}, S_{gi}) \perp C_{gi}$. We propose to use a nonparametric kernel Nelson-Aalen estimator to estimate $\psi_A(t\mid s,t_0)$ as $\widehat \psi_A(t\mid s,t_0)  = \exp\{-\widehat{\Lambda}_A(t\mid s,t_0) \}$, where
$$\widehat{\Lambda}_A(t\mid s,t_0) = \int_{t_0}^t \frac{\sum_{i=1}^{n_A} I(X_{Ai}>t_0) K_h\{\gamma(S_{Ai}) - \gamma(s)\}dN_{Ai}(z)}{\sum_{i=1}^{n_A}  I(X_{Ai}>t_0) K_h\{\gamma(S_{Ai}) - \gamma(s)\} Y_{Ai}(z)},$$
is a consistent estimate of $\Lambda_A(t\mid s,t_0 ) =  -\log [\psi_A(t\mid s,t_0)],$ $Y_{Ai}(t) = I(X_{Ai} \geq t)$,  $N_{Ai}(t) = I(X_{Ai} \leq t) \delta_i,  K(\cdot)$ is a smooth symmetric density function, $K_h(x) = K(x/h)/h$, $\gamma(\cdot)$ is a given monotone transformation function, and $h$ is a specified bandwidth. To obtain an appropriate $h$, we require the standard undersmoothing assumption of $h=O(n_A^{-u})$ with $u \in (1/4,1/2)$ in order to eliminate the impact of the bias of the conditional survival function on the resulting estimator. We first use the bandwidth selection procedure given by  \cite{scott1992multivariate} to obtain $h_{opt}$; and then we let $h = h_{opt}n_A^{-c_0}$  for some  $c_0 \in (1/20, 3/10)$ to ensure the desired rate for $h$. In all numerical examples, we chose {$c_0 = 0.11.$}  { Since $F_B(s \mid t_0) = P(S_{Bi} \le s \mid X_{Bi} > t_0)$, we empirically estimate $F_B(s \mid t_0)$ using all subjects with $X_{Bi} > t_0$ as
$$
\Fhat_B(s \mid t_0) = \frac{\sum_{i=1}^{n_B} I(S_{Bi} \le s, X_{Bi} > t_0)}{\sum_{i=1}^{n_B} I(X_{Bi} > t_0)}.
$$
Subsequently, we may construct an estimator for $\Delta_{S}(t,t_0)$ as}
$$\widehat{\Delta}_S(t,t_0) = n_B^{-1} \sum_{i=1}^{n_B} \left[\widehat{\psi}_A(t\mid S_{Bi},t_0) \frac{I(X_{Bi} > t_0)}{\widehat{W}^C_B(t_0)}  -   \frac{I(X_{Bi} > t)}{\widehat{W}^C_B(t)}\right] $$
and $\widehat{R}_S(t,t_0) =1- \widehat{\Delta}_S(t,t_0)/\widehat{\Delta}(t).$ In Appendix B in the Supplementary Materials, we show that under mild regularity conditions $\widehat{\Delta}_S(t,t_0)$ is a consistent estimator of $\Delta_S(t,t_0)$ and that as $n_A, n_B \rightarrow \infty$,
$$\sqrt{n}\left(\begin{array}{c}\widehat{\Delta}_S(t,t_0)-\Delta_S(t,t_0) \\ \widehat{\Delta}(t)-\Delta(t)\end{array}\right)\rightarrow N(0, \Sigma_\Delta),$$
where $n=n_A+n_B$. It then follow that $\widehat{R}_S(t,t_0)$ is a consistent estimator of $R_S(t,t_0)$ and, by the delta method,
$\sqrt{n}\{\widehat{R}_S(t,t_0)-R_S(t,t_0)\}$ converges weakly to a mean zero normal distribution with a variance of $\sigma_R^2.$

\subsection{Augmentation for improved efficiency using baseline covariates \label{aug}}
Recent work has shown that augmentation can lead to improvements in efficiency by taking advantage of the association between baseline information, $\bZ$, and the primary outcome (\cite*{TianCai12,garcia2011efficiency,zhang2008improving}). To investigate whether it is possible to gain efficiency through augmentation in this setting we propose the augmented estimates:
\begin{eqnarray}
\left(\begin{array}{c} \widehat{\Delta}(t)^{AUG} \\ \widehat{\Delta}_S(t,t_0)^{AUG}\end{array}\right)=\left(\begin{array}{c} \widehat{\Delta} (t) \\ \widehat{\Delta}_S(t, t_0)\end{array}\right)+{\cal A}\left\{n_A^{-1}\sum_{i=1}^{n_A}h(Z_{Ai})-n_B^{-1}\sum_{i=1}^{n_B}h(Z_{Bi})\right\} \label{def-delta-aug}
 \end{eqnarray}
 and
 $$\widehat{R}(t,t_0)^{AUG}=1-\frac{\widehat{\Delta}_S(t,t_0)^{AUG}}{\widehat{\Delta}(t)^{AUG}}$$
 where $Z_{gi}, i=1, 2, \cdots, n_g$ are i.i.d. random vectors of baseline covariates from treatment group $g$ and $h(\cdot)$ is a basis transformation given a priori.  Due to treatment randomization, $n_A^{-1}\sum_{i=1}^{n_A}h(Z_{Ai})-n_B^{-1}\sum_{i=1}^{n_B}h(Z_{Bi})$ converges to zero in probability as the sample size goes to infinity and thus the augmented estimator converges to the same limit as the original counterparts.  We propose to select ${\cal A}$ such that the variance of $(\widehat{\Delta}(t)^{AUG}, \widehat{\Delta}_S(t,t_0) ^{AUG})'$ is minimized.  That is,  ${\cal A} = (\Xi_{12}) ( \Xi_{22} ) ^{-1}$ where
 \begin{eqnarray*}
 \Xi_{12} &=& \mbox{cov} \left \{\left(\begin{array}{c} \widehat{\Delta}(t) \\ \widehat{\Delta}_S(t,t_0)\end{array}\right), n_A^{-1}\sum_{i=1}^{n_A}h(Z_{Ai})-n_B^{-1}\sum_{i=1}^{n_B}h(Z_{Bi})\right \}, \\
  \Xi_{22} &=& \mbox{var} \left \{n_A^{-1}\sum_{i=1}^{n_A}h(Z_{Ai})-n_B^{-1}\sum_{i=1}^{n_B}h(Z_{Bi})\right \}
\end{eqnarray*}
and thus we can obtain $\widehat{\Delta}(t)^{AUG}$ by replacing ${\cal A}$ with a consistent estimator, $\widehat{{\cal A}}$. We approximate ${\cal A}$ using a perturbation resampling approach described in Section \ref{inference}. 
\def\Fbb{\mathbb{F}}

\subsection{Incremental value of Surrogate Marker $S$ Measurements at $t_0$}
Since our definition of $\Delta_S(t,t_0)$ considers the surrogate information as a combination of both $S$ information and $T$ information up to $t_0$, a logical inquiry would be how to assess the incremental value of the $S$ information in terms of the proportion of treatment effect explained, when added to $T$ information up to $t_0$. If the quantity $R_S(t,t_0)$ reveals that a large proportion of the treatment effect is explained by information at $t_0$, it would be important to know how much of that quantity is attributable to $S$ information. If most of the surrogacy is due to $T$ information up to $t_0$, then it may not be necessary to measure and incorporate $S$ information. { Similar to our definition of $\Delta_S(t,t_0)$ in Section \ref{causal},} we define the proportion of treatment effect explained by $T$ information up to $t_0$ only as $R_T(t,t_0) = 1-\Delta_T(t,t_0)/\Delta(t)$ where { $\Delta_T(t, t_0)$ can be obtained by
replacing $S\supA = S\supB = 1$, which leads to
\begin{align*}
\Delta_T(t, t_0) 
& =  \sum_{d_{t_0} = 0}^1 E\{  I(T \supA > t) - I( T\supB>t) \mid  I(T\supA > t_0) = I(T\supB > t_0) = d_{t_0} \}  \Fbb({d_{t_0}}) \\
& = \Fbb(1) \{P(T \supA > t \mid T\supA > t_0) - P(T\supB > t \mid T\supB > t_0)\}
\end{align*}
where $\Fbb(d) = P(\Isc = d)$ is the probability mass function for a binary random variable $\Isc$.}

As with $\Delta_S(t,t_0)$, the choice of $\Fbb$ depends on the specific context. We will continue to assume, without loss of generality,
that treatment B is a placebo group and thus it is reasonable to consider { $\Fbb(1) = P(T\supB > t_0)$} as the reference distribution. It follows that
$$\Delta_T(t, t_0) = P(T^{(B)}>t_0)P(T^{(A)}>t\mid T^{(A)}>t_0)-P(T^{(B)}>t).$$  Although one would generally expect that the proportion of treatment effect explained by both $S$ and $T$ information up to $t_0$ to be at least as big as the proportion of treatment effect explained by $T$ information up to $t_0$ alone (i.e. $\Delta(t) - \Delta_T(t,t_0) \le  \Delta(t) - \Delta_S(t,t_0)$ implying $ \Delta_T(t,t_0) \ge  \Delta_S(t,t_0)$), this is only guaranteed to hold under certain conditions. Specifically, we note that
\begin{eqnarray*}
 \Delta_T(t,t_0) - \Delta_S(t,t_0) &=& P(T^{(B)}>t_0) \int \psi_A(t \mid s, t_0) d\left\{F_A(s\mid t_0)-F_B(s \mid t_0) \right \}
\end{eqnarray*}
and therefore $ \Delta_T(t,t_0) - \Delta_S(t,t_0) \ge 0$ if and only if $$\int \psi_A(t \mid s, t_0) d\left\{F_A(s\mid t_0)-F_B(s \mid t_0) \right \}\ge 0.$$ { Sufficient conditions for the inequality above are (C1) and $$P(S\supA>s|T\supA>t_0)>P(S\supB>s|T\supB>t_0) \mbox{ for all } s.$$ } {  These conditions are also required to ensure that we are not in a situation known as the surrogate paradox \cite{vanderweele2013surrogate}.} { When these conditions hold, it would be of interest to quantify the surrogacy incremental value of $S$ information as}
\begin{equation}
IV_S(t,t_0) = \frac{\Delta_T(t,t_0) - \Delta_S(t,t_0)}{\Delta (t)}.\label{IV}
\end{equation}

{To estimate $R_T(t_0)$, we may employ the IPW estimator} $\widehat{R}_T(t,t_0) = 1-\widehat{\Delta}_T(t,t_0)/\widehat{\Delta}(t)$ where
$\widehat{\Delta}_T(t,t_0) =  \widehat{\phi}_B(t_0)\widehat{\phi}_A(t)/\widehat{\phi}_A(t_0) - \widehat{\phi}_B(t)$ and $\widehat{\phi}_g(u) = n_g^{-1} \sum_{i=1}^{n_g} \frac{I(X_{gi}>u)}{\widehat{W}^C_g(u)}$ for $g=A,B$. { Subsequently, we may construct a plug-in estimator for $IV_S(t,t_0)$  by replacing} $\Delta(t),\Delta_S(t,t_0),$ and  $\Delta_T(t,t_0)$ with $\widehat{\Delta}(t),\widehat{\Delta}_S(t,t_0),$ and  $\widehat{\Delta}_T(t,t_0),$ respectively.



\section{Inference and Variance Estimation using Perturbation-Resampling \label{inference}}

We propose to estimate the variability of our proposed estimators and construct confidence intervals using a perturbation-resampling method to approximate the distribution of the estimators. Specifically, let $\left \{ \mathbf{V}^{(b)} = (V_{A1}^{(b)}, ...V_{An_A}^{(b)}, V_{B1}^{(b)}, ...V_{Bn_B}^{(b)})\trans, b=1,....,D \right \}$ be $n \times D$ independent copies of a positive random variables $V$ from a known distribution with unit mean and unit variance, such as the standard exponential distribution. Let

$$\widehat{\Delta}^{(b)} (t) = \frac{  \sum_{i=1}^{n_A} V_{Ai}^{(b)} I(X_{Ai}>t)}{ \sum_{i=1}^{n_A} V_{Ai}^{(b)} \widehat{W}_A^{C(b)}(t)}  -\frac{  \sum_{i=1}^{n_B} V_{Bi}^{(b)} I(X_{Bi}>t)}{ \sum_{i=1}^{n_B} V_{Bi}^{(b)} \widehat{W}_B^{C(b)}(t)},$$

$$\widehat{\Lambda}^{(b)}_A(t\mid s, t_0) = \int_{t_0}^t \frac{\sum_{i=1}^{n_A} V_{Ai}^{(b)}  I(X_{Ai}>t_0) K_h\{\gamma(S_{Ai}) - \gamma(s)\}dN_i(z)}{\sum_{i=1}^{n_A}  V_{Ai}^{(b)} I(X_{Ai}>t_0) K_h\{\gamma(S_{Ai}) - \gamma(s)\} Y_i(z)},$$

$$\widehat{\Delta}^{(b)} _S(t,t_0)  = \frac{ \sum_{i=1}^{n_B} V_{Bi}^{(b)}  \widehat{\psi}^{(b)}_ A(t\mid S_{Bi},t_0)  I(X_{Bi} > t_0)}{ \sum_{i=1}^{n_B} V_{Bi}^{(b)}  \widehat{W}^{C(b)}_B(t_0)}  -  \frac{ \sum_{i=1}^{n_B}V_{Bi}^{(b)} I(X_{Bi}>t)}{ \sum_{i=1}^{n_B} V_{Bi}^{(b)}  \widehat{W}^{C(b)}_B(t)},$$

and $$\widehat{R}^{(b)} _S(t,t_0)  = 1-\widehat{\Delta}^{(b)} _S(t, t_0)/\widehat{\Delta}^{(b)}(t)$$

\noindent where $\widehat{\psi}_A^{(b)} (t\mid s, t_0)  = \exp\{-\widehat{\Lambda}_A^{(b)} (t\mid s, t_0) \}$ and $\widehat{W}_g^{C(b)}(\cdot)$ is the Kaplan-Meier estimator of survival for censoring with weights $V_{gi}^{(b)}$ for $g=A, B$. Then one can estimate the distribution of
\begin{equation}
\left(\begin{array}{c}\widehat{\Delta}_S(t,t_0)-\Delta_S(t,t_0)\\ \widehat{\Delta}(t)-\Delta(t) \end{array}\right) \label{var1}
\end{equation}
by the empirical distribution of
\begin{equation}
\left(\begin{array}{c}\widehat{\Delta}^{(b)}_S(t,t_0)-\widehat{\Delta}_S(t,t_0) \\ \widehat{\Delta}^{(b)} (t)-\widehat{\Delta}(t)\end{array}\right), b = 1,...,D.\label{var2}
\end{equation}
That is, one can approximate the variance of (\ref{var1}) with the empirical variance of (\ref{var2}), denoted as $\widehat{\Sigma}$. To construct a $100(1-\alpha)\%$ confidence interval for $R_S(t,t_0)$, one can calculate the $100(\alpha/2)^{\mbox{th}}$ and $100(1-\alpha/2)^{\mbox{th}}$ empirical percentile of $\widehat{R}^{(b)} _S(t,t_0)$ or estimate the variance of $\widehat R_S(t,t_0)-R_S(t,t_0)$ by the empirical variance of $\widehat{R}^{(b)} _S(t,t_0)-\widehat{R}_S(t,t_0)$ and construct the corresponding Wald-type confidence interval. An alternative is to employ Fieller's method for making inference on the ratio of two parameters (\cite*{fieller1954some,fieller1940biological}) and obtain the $100(1-\alpha)$\% confidence interval for $R_S(t,t_0)$ as
$$\left\{r:  \frac{\{\widehat{\Delta}_S(t,t_0)-(1-r)\widehat{\Delta}(t,t_0)\}^2}{\hat{\sigma}_{11}-2(1-r)\hat\sigma_{12}+(1-r)^2\hat\sigma_{22}} \le c_{\alpha}\right\},$$
where $\widehat{\Sigma}=(\hat\sigma_{ij})_{1\le i,j\le 2}$ and $c_\alpha$ is the $100\times(1-\alpha)$th percentile of
$$\left\{\frac{[\widehat{\Delta}^{(b)}_S (t,t_0)-\{1-\widehat R_S(t,t_0)\}\widehat{\Delta}^{(b)} (t,t_0)]^2}{\hat{\sigma}_{11}-2\{1-\widehat R_S(t,t_0)\}\hat\sigma_{12}+\{1-\widehat R_S(t,t_0)\}^2\hat\sigma_{22}}, b=1, \cdots, D\right\}.$$ The theoretical justification for the perturbation-resampling procedure is provided in Appendix C in the Supplementary Materials.

{ The perturbed samples can also be used to construct the augmented estimators $\widehat{\Delta}(t)^{AUG}$ and  $\widehat{\Delta}_S(t,t_0)^{AUG}$, defined
in (\ref{def-delta-aug}), by replacing ${\cal A}$ with
$\widehat{{\cal A}} = (\widehat{\Xi}_{12} ) ( \widehat{\Xi}_{22} ) ^{-1}$, where $\widehat{\Xi}_{12}$ is the empirical covariance of
  $$\left \{\left(\begin{array}{c} \widehat{\Delta}^{(b)} (t) \\ \widehat{\Delta}^{(b)} _S(t,t_0)\end{array}\right),
  n_A^{-1}\sum_{i=1}^{n_A}  V_{Ai}^{(b)} h(Z_{Ai})-n_B^{-1}\sum_{i=1}^{n_B}  V_{Bi}^{(b)} h(Z_{Bi})\right \}, b=1,...,D$$
and $\widehat{\Xi}_{22}$ is the empirical variance of

$$ \left \{n_A^{-1}\sum_{i=1}^{n_A}  V_{Ai}^{(b)} h(Z_{Ai})-n_B^{-1}\sum_{i=1}^{n_B}  V_{Bi}^{(b)} h(Z_{Bi})\right \}, b=1,...,D.$$
The estimator $\widehat{R}(t,t_0)^{AUG}$ can be constructed accordingly.}


\section{Numerical Studies}
\subsection{Simulation Studies \label{sims}}
We conducted simulation studies under two main settings to assess the performance and validity of our proposed estimators and inference procedures. In both settings, data were generated such that individuals may experience the primary outcome or be censored before $t_0$ and thus,  $S$ is only measured on individuals still under observation at $t_0$. {  Within each setting we examined results where $n_A = n_B = 1000$ and $n_A = n_B = 400$. } Throughout,  we use a normal density kernel, $t=1$, $t_0 = 0.5$, and the results summarize 1000 replications. For all estimates, we estimate variance using our proposed perturbation approach and construct confidence intervals using the normal approximation, quantiles of the perturbed values and Fieller's method (for $R_S(t,t_0)$ only).

In the first simulation setting, Setting (i), data were generated as:
\begin{eqnarray*}
S^{(A)} &\sim& \mbox{Gamma}(\mbox{shape = 2, scale = 2})\\
S^{(B)} &\sim& \mbox{Gamma}(\mbox{shape = 9, scale = 0.5})\\
T^{(A)} \mid S^{(A)}, Z^{(A)} &=& -\mbox{log}(1-Z^{(A)})/(0.2\times S^{(A)}) \\
T^{(B)} \mid S^{(B)}, Z^{(B)}  &=& -\mbox{log}(1-Z^{(B)})/(0.2+0.22\times S^{(B)})
\end{eqnarray*}
\noindent and $Z^{(A)}$ and $Z^{(B)}$ were generated from a $\mbox{N}(0,1)$  distribution and censoring in both groups was simulated as $C\sim Exp(0.5)$, where $S^{(g)}$ is only observable if $T^{(g)} > t_0$ and $C>t_0$. In this setting, $\Delta(t) = 0.19$, $\Delta_S(t,t_0)=0.05$, $R_S(t,t_0)=0.75$, $P(T\supA > t) =0.51$, $P(T\supB > t) =0.32$, $P(T\supA > t_0) =0.69$, $P(T\supB > t_0) =0.56$, $E(S\supA | T \supA >t_0) = 3.35$, $E(S\supB | T \supB >t_0) = 4.27$, and 29\% and 25\% of individuals in treatment group A and treatment group B are censored before $t$, respectively. The top portion of Table \ref{table-conditional} shows the results from this setting when $n_A = n_B=1000$. These results show that in finite samples the proposed estimates have very small bias and adequate coverage, the standard error estimates obtained from the perturbation-resampling procedure are close to the average standard error estimates, and augmentation provides some efficiency gain.

In the second simulation setting (ii),  data were generated as:
\begin{eqnarray*}
S^{(A)} &\sim& \mbox{Gamma}(\mbox{shape = 2, scale = 2})\\
S^{(B)} &\sim& \mbox{Gamma}(\mbox{shape = 9, scale = 0.5})\\
T^{(A)} \mid S^{(A)}, Z^{(A)} &=& \exp\{ 0.5*S \supA +1.5 Z\supA + \epsilon_1\}, \quad \epsilon_1 \sim \mbox{N}(0.5,1) \\
T^{(B)} \mid S^{(B)}, Z^{(B)} &=& \exp\{ 0.1*S \supB +1.5 Z\supB + \epsilon_2\}, \quad \epsilon_1 \sim \mbox{N}(0,1) 
\end{eqnarray*}
\noindent and $Z^{(A)}$ and $Z^{(B)}$ were generated from a $\mbox{N}(0,1)$  distribution and censoring in both groups was simulated as $C= B*e_1 + (1-B)*e_2$, where $B \sim \mbox{Bernoulli}(0.5), e_1 \sim \mbox{Exp}(0.5), e_2 \sim \mbox{Exp}(0.3)$. In this setting, $\Delta(t) = 0.27$, $\Delta_S(t,t_0)=0.11$, $R_S(t,t_0)=0.60$, $P(T\supA > t) =0.87$, $P(T\supB > t) =0.61$, $P(T\supA > t_0) =0.93$, $P(T\supB > t_0) =0.75$, $E(S\supA | T \supA >t_0) = 4.16$, $E(S\supB | T \supB >t_0) = 4.56$, 30\% and 25\% of individuals in treatment group A and treatment group B are censored before $t$, respectively. The bottom portion of Table \ref{table-conditional} shows the results from this setting when $n_A=n_B=1000$. Similar to setting (i), these results show that our proposed estimate performs well in finite samples; specifically, the bias is very small, the coverage is adequate,  the standard error estimates obtained from the perturbation-resampling procedure are close to the average standard error estimates, and augmentation provides some efficiency gain. For comparison, the estimate of \cite{lin1997estimating} in this setting was -0.13 using Cox models and -0.40 using AFT models. In addition, we simulated data in this same setting with the exception that censoring in both groups was simulated as $C = \exp\{N(4,1)\}$; with this change, the estimate of \cite{lin1997estimating}  when $n_A = n_B = 1000$  was -0.06 using Cox models and -0.76 using AFT models while the estimates from our proposed procedure look almost identical to those shown in Table \ref{table-conditional}. The fact that this model-based approach a) provides a negative estimate of the proportion of treatment effect explained by the surrogate and b) provides two rather different estimates when only the censoring distribution is changed demonstrates the advantage of utilizing a method that does not require strict modeling assumptions in settings where they may not hold. 

Table \ref{table-conditional400} shows results for both settings when $n_A = n_B = 400$. With a smaller sample size, we recommend using the median absolute variance when calculating the empirical variance of the perturbed quantities (see Section  \ref{inference}) to guard against the possibility that an outlier perturbed sample will dramatically influence the variance estimates. These results show that the proposed procedure still performs reasonably well with smaller sample sizes. As expected, the variance is larger compared to the large sample size setting but the bias is similar and the coverage is adequate.

\subsection{Example \label{example}}
We illustrate our proposed procedures using data from the Diabetes Prevention Program (DPP), a randomized clinical trial designed to investigate the efficacy of various treatments on the prevention of type 2 diabetes in high-risk adults. At randomization, participants were randomly assigned to one of four groups: metformin, troglitazone, lifestyle intervention or placebo. The troglitazone arm of the study was discontinued due to medication toxicity. The primary endpoint was time to diabetes as defined by the protocol at the time of the visit: fasting glucose $\geq$ 140 mg/dL (for visits  through 6/23/1997, $\geq$ 126 mg/dL for visits on or after 6/24/2007) or 2-hour post challenge glucose $\geq$ 200 mg/dL.  DPP results showed that both lifestyle intervention and metformin prevented or delayed development of type 2 diabetes in high risk adults (\cite*{diabetes1999diabetes,DPPOS_NEJM}).

For this illustration, we focus on the comparison of the lifestyle intervention group (N=1024) vs. placebo (N=1030) and we aim to examine the proportion of treatment effect explained by two potential surrogate markers:  change in log-transformed  hemoglobin A1c (HBA1C) from baseline to $t_0=1$ year and change in fasting plasma glucose from baseline to $t_0$. We define the treatment effect, $\Delta(t)$, as the difference in diabetes prevalence at $t= 3$ years after randomization. Individuals who die before 3 years are censored at the last time point where glucose was measured. The estimated probability of not developing diabetes by $t=3$ years was 0.86 in the lifestyle intervention group and 0.71 in the placebo group; therefore the treatment effect $\widehat{\Delta}(t) = 0.86-0.71 = 0.15$. Since we define surrogate information at $t_0$ as including diabetes incidence at $t_0$, it is interesting to note that 3.7\% and 11.5\% of participants in the lifestyle intervention group and placebo group were diagnosed with diabetes before $t_0=1$ year, respectively. 

Results from estimating the proportion of treatment effect explained by each surrogate are shown in Table \ref{table-example}. Using our proposed procedure, the estimated residual treatment effect, $\widehat{\Delta}_S(t,t_0)$, is 0.077 when the surrogate information at 1 year post-baseline consists of information about the change in HBA1C from baseline to 1 year and diabetes incidence up to 1 year and the proportion of treatment effect explained by this surrogate information, $R_S(t,t_0)$, is 48.2\%. Examining change in fasting plasma glucose, the estimated residual treatment effect is 0.046 when the surrogate information at 1 year post-baseline consists of information about the change in fasting plasma glucose from baseline to 1 year and diabetes incidence up to 1 year and the proportion of treatment effect explained by this surrogate information is 68.7\%. To determine the incremental value of the information about change in hemoglobin A1c and fasting plasma glucose, we examined the proportion of treatment effect explained by diabetes incidence information only up to 1 year post-baseline which was estimated to be 47.8\%. Therefore, the incremental value of change in HBA1C was negligible, while the incremental value of change in fasting plasma glucose was 21.0\% (SE= 5.9\%). Our application of the proposed procedures to examine surrogate markers shows that fasting plasma glucose appears to capture more of the treatment effect than hemoglobin A1c, particularly when considered in terms of incremental value when added to diabetes incidence information at 1 year post-baseline.

To examine whether efficiency could be gained through augmentation, we also calculated our proposed augmented estimates using the available baseline covariates: age group (less than 40, 40-44,45-49,50-54,55-59,60-64,65 and older), body mass index category (km/m$^2$ units, $<26$, $\geq 26$ to $< 28$, $\geq 28$ to $< 30$, $\geq 30$ to $< 32$, $\geq 32$ to $< 34$, $\geq 34$ to $< 36$, $\geq 36$ to $< 38$, $\geq 38$ to $< 40$, $\geq 40$ to $< 42$, and $\geq 42$), self-reported race/ethnicity (Caucasian, African American, Hispanic, other), and gender. The resulting estimates for change in HBA1C were $\widehat{\Delta}(t) ^{AUG} =0.15 (SE = 0.019)$, $\widehat{\Delta}_S(t,t_0) ^{AUG} = 0.078(SE = 0.017)$, and $\widehat{R}_S(t,t_0) ^{AUG} = 0.48 (SE = 0.10)$.  The resulting estimates for change in fasting plasma glucose were $\widehat{\Delta}(t) ^{AUG} =0.15 (SE = 0.019)$, $\widehat{\Delta}_S(t,t_0) ^{AUG} = 0.05 (SE = 0.02)$, and $\widehat{R}_S(t,t_0) ^{AUG} = 0.69 (SE = 0.10)$. That is, in this particular example, the use of baseline covariates through augmentation leads to little to no improvement in efficiency. Our application of the proposed procedures to examine surrogate markers shows that fasting plasma glucose appears to capture more of the treatment effect than HBA1C, particularly when considered in terms of incremental value when added to diabetes incidence information at 1 year post-baseline.

\section{Discussion}
The identification and validation of surrogate markers is an important and challenging area of research. Valid surrogate markers that could be used to replace the primary outcome or used in combination with primary outcome information have the potential to lead to gains in efficiency in terms of design, implementation, estimation and testing. In this paper we have proposed a novel { model-free} framework for quantifying the proportion of treatment effect explained by surrogate information collected up to a specified time in the survival setting and a robust nonparametric procedure for making inference. {Our proposed methods also have the advantage of allowing
the surrogate marker $S$ to be not observable at time $t_0$.}

While we have defined the treatment effect quantity of interest as the difference in survival at time $t$, our proposed definition and estimation procedure can be extended to other treatment effect quantities such as the restricted mean survival time. Another option would be to define a treatment effect quantity over time. When $T$ is the time of a non-terminal primary outcome such as time until diabetes diagnosis, there are two important considerations. First, competing risks must be accounted for in estimation since death could censor the observation of the primary outcome. Specifically, one would censor an individual at the time of death and apply the procedures proposed above. Second, it may be possible for the surrogate marker to still be observed after the primary outcome occurs is individuals are still under observation in the study. However, depending on the setting, it may not be appropriate to incorporate this surrogate marker information when estimating the proportion of treatment effect because treatment decisions made after the primary outcome occurs may affect the surrogate marker measurement in ways that would make the proportion of treatment effect estimation uninterpretable.  On the other hand, if the non-terminal primary outcome is an outcome that is not easily observed and/or requires expensive, invasive or time-intensive testing to determine whether it occurred, then use of a surrogate marker that may be measured after the event occurs may be of interest.

{  
The simulation study shows that the proposed inference procedure has satisfactory empirical performance for moderate sample sizes. When the sample size becomes much smaller, these procedures which are based on asymptotic normality approximations would still lead to reliable inference for $\Delta_S(t, t_0)$ and $\Delta(t)$. However, the the asymptotical normality approximation of $R_S(t,t_0)$, which involves the ratio of $\Delta_S(t, t_0)$ and $\Delta(t)$, would likely be less reliable and the proposed inference method may not be very accurate. 

Lastly, a limitation of our proposed approach is the theoretical condition that the supports of $S^{(A)}$ and $S^{(B)}$ are equivalent.  In practice, the empirical supports may not completely overlap and some type of transformation or extrapolation of the relevant nonparametric estimators may be needed.  However, when there is  substantial non-overlap between two supports,  caution is needed in  interpreting the results. }

\section*{Acknowledgements}
Support for this research was provided by National Institutes of Health grant R21DK103118. The Diabetes Prevention Program (DPP) was conducted by the DPP Research Group and supported by the National Institute of Diabetes and Digestive and Kidney Diseases (NIDDK), the General Clinical Research Center Program, the National Institute of Child Health and Human Development (NICHD), the National Institute on Aging (NIA), the Office of Research on Women's Health, the Office of Research on Minority Health, the Centers for Disease Control and Prevention (CDC), and the American Diabetes Association. The data from the DPP were supplied by the NIDDK Central Repositories. This manuscript was not prepared under the auspices of the DPP and does not represent analyses or conclusions of the DPP Research Group, the NIDDK Central Repositories, or the NIH.

\bibliographystyle{biometri}
\bibliography{Surrogate_bib}

\begin{thebibliography}{}

\bibitem[\protect\astroncite{Buyse \&\ Molenberghs}{Buyse \&\
  Molenberghs}{1998}]{buyse1998criteria}
{\sc Buyse, M. \&\ Molenberghs, G.} (1998).
\newblock Criteria for the validation of surrogate endpoints in randomized
  experiments.
\newblock {\em Biometrics} {\bf 54}, 1014--1029.

\bibitem[\protect\astroncite{Conlon, Taylor \&\ Elliott}{Conlon
  et~al.}{2014a}]{conlon2014surrogacy2}
{\sc Conlon, A., Taylor, J. \&\ Elliott, M.} (2014a).
\newblock Surrogacy assessment using principal stratification and a {G}aussian
  copula model.
\newblock {\em Statistical {M}ethods in {M}edical {R}esearch} {\bf June 19},
  Epub ahead of print.

\bibitem[\protect\astroncite{Conlon, Taylor \&\ Elliott}{Conlon
  et~al.}{2014b}]{conlon2014surrogacy}
{\sc Conlon, A.~S., Taylor, J.~M. \&\ Elliott, M.~R.} (2014b).
\newblock Surrogacy assessment using principal stratification when surrogate
  and outcome measures are multivariate normal.
\newblock {\em Biostatistics} {\bf 15}, 266--283.

\bibitem[\protect\astroncite{{Diabetes~Prevention~Program~Group}}{{Diabetes~Prevention~Program~Group}}{1999}]{diabetes1999diabetes}
{\sc {Diabetes~Prevention~Program~Group}} (1999).
\newblock The diabetes prevention program: design and methods for a clinical
  trial in the prevention of {T}ype 2 diabetes.
\newblock {\em Diabetes {C}are} {\bf 22}, 623--634.

\bibitem[\protect\astroncite{{Diabetes~Prevention~Program~Group}}{{Diabetes~Prevention~Program~Group}}{2002}]{DPPOS_NEJM}
{\sc {Diabetes~Prevention~Program~Group}} (2002).
\newblock Reduction in the incidence of {T}ype 2 diabetes with lifestyle
  intervention or {M}etformin.
\newblock {\em New England Journal of Medicine} {\bf 346}, 393--403.
\newblock PMID: 11832527.

\bibitem[\protect\astroncite{Fieller}{Fieller}{1940}]{fieller1940biological}
{\sc Fieller, E.} (1940).
\newblock The biological standardization of insulin.
\newblock {\em Supplement to the Journal of the Royal Statistical Society} {\bf
  7}, 1--64.

\bibitem[\protect\astroncite{Fieller}{Fieller}{1954}]{fieller1954some}
{\sc Fieller, E.~C.} (1954).
\newblock Some problems in interval estimation.
\newblock {\em Journal of the Royal Statistical Society. Series B
  (Methodological)} {\bf 16}, 175--185.

\bibitem[\protect\astroncite{Frangakis \&\ Rubin}{Frangakis \&\
  Rubin}{2002}]{frangakis2002principal}
{\sc Frangakis, C.~E. \&\ Rubin, D.~B.} (2002).
\newblock Principal stratification in causal inference.
\newblock {\em Biometrics} {\bf 58}, 21--29.

\bibitem[\protect\astroncite{Freedman, Graubard \&\ Schatzkin}{Freedman
  et~al.}{1992}]{freedman1992statistical}
{\sc Freedman, L.~S., Graubard, B.~I. \&\ Schatzkin, A.} (1992).
\newblock Statistical validation of intermediate endpoints for chronic
  diseases.
\newblock {\em Statistics in {M}edicine} {\bf 11}, 167--178.

\bibitem[\protect\astroncite{Gabriel \&\ Gilbert}{Gabriel \&\
  Gilbert}{2014}]{gabriel2014evaluating}
{\sc Gabriel, E.~E. \&\ Gilbert, P.~B.} (2014).
\newblock Evaluating principal surrogate endpoints with time-to-event data
  accounting for time-varying treatment efficacy.
\newblock {\em Biostatistics} {\bf 15}, 251--265.

\bibitem[\protect\astroncite{Gabriel, Sachs \&\ Gilbert}{Gabriel
  et~al.}{2015}]{gabriel2015comparing}
{\sc Gabriel, E.~E., Sachs, M.~C. \&\ Gilbert, P.~B.} (2015).
\newblock Comparing and combining biomarkers as principle surrogates for
  time-to-event clinical endpoints.
\newblock {\em Statistics in {M}edicine} {\bf 34}, 381--395.

\bibitem[\protect\astroncite{Garcia, Ma \&\ Yin}{Garcia
  et~al.}{2011}]{garcia2011efficiency}
{\sc Garcia, T.~P., Ma, Y. \&\ Yin, G.} (2011).
\newblock Efficiency improvement in a class of survival models through
  model-free covariate incorporation.
\newblock {\em Lifetime {D}ata {A}nalysis} {\bf 17}, 552--565.

\bibitem[\protect\astroncite{Ghosh}{Ghosh}{2008}]{ghosh2008semiparametric}
{\sc Ghosh, D.} (2008).
\newblock Semiparametric inference for surrogate endpoints with bivariate
  censored data.
\newblock {\em Biometrics} {\bf 64}, 149--156.

\bibitem[\protect\astroncite{Ghosh}{Ghosh}{2009}]{ghosh2009assessing}
{\sc Ghosh, D.} (2009).
\newblock On assessing surrogacy in a single trial setting using a
  semicompeting risks paradigm.
\newblock {\em Biometrics} {\bf 65}, 521--529.

\bibitem[\protect\astroncite{Gilbert \&\ Hudgens}{Gilbert \&\
  Hudgens}{2008}]{gilbert2008evaluating}
{\sc Gilbert, P.~B. \&\ Hudgens, M.~G.} (2008).
\newblock Evaluating candidate principal surrogate endpoints.
\newblock {\em Biometrics} {\bf 64}, 1146--1154.

\bibitem[\protect\astroncite{Huang \&\ Gilbert}{Huang \&\
  Gilbert}{2011}]{huang2011comparing}
{\sc Huang, Y. \&\ Gilbert, P.~B.} (2011).
\newblock Comparing biomarkers as principal surrogate endpoints.
\newblock {\em Biometrics} {\bf 67}, 1442--1451.

\bibitem[\protect\astroncite{Lin, Fleming, DeGruttola et~al.}{Lin
  et~al.}{1997}]{lin1997estimating}
{\sc Lin, D., Fleming, T., DeGruttola, V. et~al.} (1997).
\newblock Estimating the proportion of treatment effect explained by a
  surrogate marker.
\newblock {\em Statistics in {M}edicine} {\bf 16}, 1515--1527.

\bibitem[\protect\astroncite{Parast, McDermott \&\ Tian}{Parast
  et~al.}{2016}]{parast2016}
{\sc Parast, L., McDermott, M.~M. \&\ Tian, L.} (2016).
\newblock Robust estimation of the proportion of treatment effect explained by
  surrogate marker information.
\newblock {\em Statistics in Medicine} .

\bibitem[\protect\astroncite{Prentice}{Prentice}{1989}]{prentice1989surrogate}
{\sc Prentice, R.~L.} (1989).
\newblock Surrogate endpoints in clinical trials: definition and operational
  criteria.
\newblock {\em Statistics in medicine} {\bf 8}, 431--440.

\bibitem[\protect\astroncite{Robins \&\ Greenland}{Robins \&\
  Greenland}{1992}]{robins1992identifiability}
{\sc Robins, J.~M. \&\ Greenland, S.} (1992).
\newblock Identifiability and exchangeability for direct and indirect effects.
\newblock {\em Epidemiology} pages 143--155.

\bibitem[\protect\astroncite{Scott}{Scott}{1992}]{scott1992multivariate}
{\sc Scott, D.} (1992).
\newblock {\em Multivariate density estimation}.
\newblock John Wiley \& Sons.

\bibitem[\protect\astroncite{Temple}{Temple}{1999}]{temple1999surrogate}
{\sc Temple, R.} (1999).
\newblock Are surrogate markers adequate to assess cardiovascular disease
  drugs?
\newblock {\em Jama} {\bf 282}, 790--795.

\bibitem[\protect\astroncite{Tian, Cai, Zhao \&\ Wei}{Tian
  et~al.}{2012}]{TianCai12}
{\sc Tian, L., Cai, T., Zhao, L. \&\ Wei, L.} (2012).
\newblock On the covariate-adjusted estimation for an overall treatment
  difference with data from a randomized comparative clinical trial.
\newblock {\em Biostatistics} {\bf 13}, 256--273.

\bibitem[\protect\astroncite{VanderWeele}{VanderWeele}{2013}]{vanderweele2013surrogate}
{\sc VanderWeele, T.~J.} (2013).
\newblock Surrogate measures and consistent surrogates.
\newblock {\em Biometrics} {\bf 69}, 561--565.

\bibitem[\protect\astroncite{Wang \&\ Taylor}{Wang \&\
  Taylor}{2002}]{wang2002measure}
{\sc Wang, Y. \&\ Taylor, J.~M.} (2002).
\newblock A measure of the proportion of treatment effect explained by a
  surrogate marker.
\newblock {\em Biometrics} {\bf 58}, 803--812.

\bibitem[\protect\astroncite{Zhang, Tsiatis \&\ Davidian}{Zhang
  et~al.}{2008}]{zhang2008improving}
{\sc Zhang, M., Tsiatis, A.~A. \&\ Davidian, M.} (2008).
\newblock Improving efficiency of inferences in randomized clinical trials
  using auxiliary covariates.
\newblock {\em Biometrics} {\bf 64}, 707--715.

\end{thebibliography}

\clearpage

\begin{landscape}
\begin{table}

\begin{center}
\begin{tabular}{|l|c|c|c||c|c|c|} \hline
\multicolumn{1}{|l|}{}&\multicolumn{6}{c|}{Setting (i)}\\ \hline
\multicolumn{1}{|l|}{}&\multicolumn{1}{c|}{$\widehat{\Delta}(t)$}&\multicolumn{1}{c|}{$\widehat{\Delta}_S(t,t_0) $}&\multicolumn{1}{c||}{$\widehat{R}_S(t,t_0) $}&\multicolumn{1}{c|}{$\widehat{\Delta}(t) ^{AUG}$}&\multicolumn{1}{c|}{$\widehat{\Delta}_S(t,t_0) ^{AUG}$}&\multicolumn{1}{c|}{$\widehat{R}_S(t,t_0) ^{AUG}$}\\ \hline
Bias&-0.0002     &0.0020      &-0.0045    &0.0000          &0.0021     &-0.0051    \\ 
ESE&0.0254     &0.0215     &0.0962     &0.0231     &0.0213     &0.0952     \\ 
ASE&0.0249     &0.0210      &0.0988     &0.0237     &0.0208     &0.0972     \\ 
MSE&0.0006     &0.0005      &0.0093     &0.0005      &0.0005      &0.0091     \\ 
Coverage (normal)&0.951      &0.945      &0.959      &0.958      &0.947      &0.955      \\ 
Coverage (quantile)&0.945      &0.943      &0.942      &0.953      &0.944      &0.941      \\ 
Coverage (Fieller)&--&--&0.952      &--&--&0.953      \\ 
\hline
\multicolumn{1}{|l|}{}&\multicolumn{6}{c|}{Setting (ii)}\\ \hline
\multicolumn{1}{|l|}{}&\multicolumn{1}{c|}{$\widehat{\Delta}(t)$}&\multicolumn{1}{c|}{$\widehat{\Delta}_S(t,t_0) $}&\multicolumn{1}{c||}{$\widehat{R}_S(t,t_0) $}&\multicolumn{1}{c|}{$\widehat{\Delta}(t) ^{AUG}$}&\multicolumn{1}{c|}{$\widehat{\Delta}_S(t,t_0) ^{AUG}$}&\multicolumn{1}{c|}{$\widehat{R}_S(t,t_0) ^{AUG}$}\\ \hline
Bias&0.0017     &0.0017     &-0.0041    &0.0013     &0.0017     &-0.0043    \\ 
ESE&0.0206     &0.0140      &0.0439     &0.0176     &0.0139     &0.0430      \\ 
ASE&0.0208     &0.0141     &0.0436     &0.0178     &0.0139     &0.0429     \\ 
MSE&0.0004      &0.0002      &0.0019     &0.0003      &0.0002      &0.0019     \\ 
Coverage (normal)&0.948      &0.954      &0.954      &0.953      &0.949      &0.947      \\ 
Coverage (quantile)&0.943      &0.946      &0.948      &0.949      &0.942      &0.942      \\ 
Coverage (Fieller)&--&--&0.953      &--&--&0.947      \\ 
\hline
\end{tabular}
\vspace{3mm}
\end{center}
\caption{Performance of the proposed estimates, the estimated treatment effect, $\widehat{\Delta}(t)$, the estimated residual treatment effect,  $\widehat{\Delta}_S(t,t_0) $, the estimated proportion of treatment effect explained by surrogate information at $t_0$, $\widehat{R}_S(t,t_0) $, the estimated augmented treatment effect, $\widehat{\Delta}(t) ^{AUG}$, the estimated augmented residual treatment effect,  $\widehat{\Delta}_S(t,t_0) ^{AUG} $, and the estimated augmented proportion of treatment effect explained by surrogate information at $t_0$, $\widehat{R}_S(t,t_0) ^{AUG} $, in Settings (i) (top portion) and (ii) (bottom portion) when $n_A = n_B =1000$ in each group; the empirical standard error (ESE), average standard error (ASE), mean squared error (MSE) and coverage of the 95\% confidence intervals are shown for confidence intervals based on a normal approximation, a quantile-based calculation, and Fieller's  method; the proposed perturbation-resampling procedure is used for variance estimation \label{table-conditional}}
\end{table}

\begin{table}

\begin{center}
\begin{tabular}{|l|c|c|c||c|c|c|} \hline
\multicolumn{1}{|l|}{}&\multicolumn{6}{c|}{Setting (i)}\\ \hline
\multicolumn{1}{|l|}{}&\multicolumn{1}{c|}{$\widehat{\Delta}(t)$}&\multicolumn{1}{c|}{$\widehat{\Delta}_S(t,t_0) $}&\multicolumn{1}{c||}{$\widehat{R}_S(t,t_0) $}&\multicolumn{1}{c|}{$\widehat{\Delta}(t) ^{AUG}$}&\multicolumn{1}{c|}{$\widehat{\Delta}_S(t,t_0) ^{AUG}$}&\multicolumn{1}{c|}{$\widehat{R}_S(t,t_0) ^{AUG}$}\\ \hline
Bias&-0.0009     &0.0034     &-0.0049    &-0.0006     &0.0036     &-0.0062    \\ 
ESE&0.0398     &0.0329     &0.1607     &0.0362     &0.0325     &0.1552     \\ 
ASE&0.0395     &0.0326     &0.1550      &0.0375     &0.0324     &0.1517     \\ 
MSE&0.0016     &0.0011     &0.0258     &0.0013     &0.0011     &0.0241     \\ 
Coverage (normal)&0.944      &0.946      &0.95       &0.954      &0.95       &0.95       \\ 
Coverage (quantile)&0.944      &0.944      &0.942      &0.952      &0.944      &0.942      \\ 
Coverage (Fieller)&--&--&0.944      &--&--&0.947      \\ 
\hline
\multicolumn{1}{|l|}{}&\multicolumn{6}{c|}{Setting (ii)}\\ \hline
\multicolumn{1}{|l|}{}&\multicolumn{1}{c|}{$\widehat{\Delta}(t)$}&\multicolumn{1}{c|}{$\widehat{\Delta}_S(t,t_0) $}&\multicolumn{1}{c||}{$\widehat{R}_S(t,t_0) $}&\multicolumn{1}{c|}{$\widehat{\Delta}(t) ^{AUG}$}&\multicolumn{1}{c|}{$\widehat{\Delta}_S(t,t_0) ^{AUG}$}&\multicolumn{1}{c|}{$\widehat{R}_S(t,t_0) ^{AUG}$}\\ \hline
Bias&0.0025     &0.001      &-0.0006     &0.0014     &0.0011     &-0.0012    \\ 
ESE&0.0470      &0.0315     &0.0988     &0.0405     &0.0310      &0.0965     \\ 
ASE&0.0465     &0.0313     &0.1003     &0.0400       &0.0309     &0.0973     \\ 
MSE&0.0022     &0.0010      &0.0097     &0.0016     &0.001      &0.0093     \\ 
Coverage (normal)&0.949      &0.949      &0.960       &0.943      &0.953      &0.959      \\ 
Coverage (quantile)&0.947      &0.951      &0.953      &0.939      &0.952      &0.947      \\ 
Coverage (Fieller)&--&--&0.955      &--&--&0.955      \\ 
\hline
\end{tabular}
\vspace{3mm}
\end{center}
\caption{Performance of the proposed estimates, the estimated treatment effect, $\widehat{\Delta}(t)$, the estimated residual treatment effect,  $\widehat{\Delta}_S(t,t_0) $, the estimated proportion of treatment effect explained by surrogate information at $t_0$, $\widehat{R}_S(t,t_0) $, the estimated augmented treatment effect, $\widehat{\Delta}(t) ^{AUG}$, the estimated augmented residual treatment effect,  $\widehat{\Delta}_S(t,t_0) ^{AUG} $, and the estimated augmented proportion of treatment effect explained by surrogate information at $t_0$, $\widehat{R}_S(t,t_0) ^{AUG} $, in Settings (i) (top portion) and (ii) (bottom portion) when $n_A = n_B =400$ in each group; the empirical standard error (ESE), average standard error (ASE), mean squared error (MSE) and coverage of the 95\% confidence intervals are shown for confidence intervals based on a normal approximation, a quantile-based calculation, and Fieller's  method; the proposed perturbation-resampling procedure is used for variance estimation \label{table-conditional400}}
\end{table}

\end{landscape}

\clearpage

\begin{table}[hptb]
\begin{center}
\begin{tabular}{|l|c|c|c|c|c|c|} \hline
\multicolumn{1}{|l|}{}&\multicolumn{6}{c|}{Difference in HBA1C from baseline to $t_0$}\\ \hline
\multicolumn{1}{|l|}{}&\multicolumn{1}{c|}{$\Delta(t)$}&\multicolumn{1}{c|}{$\Delta_S(t,t_0)$}&\multicolumn{1}{c|}{$R_S(t,t_0)$}&\multicolumn{1}{c|}{$\Delta_T(t,t_0)$}&\multicolumn{1}{c|}{$R_T(t,t_0)$}&\multicolumn{1}{c|}{$IV_S(t,t_0)$}\\ \hline
Estimate&0.1483      &0.0769      &0.4815      &0.0774      &0.4779      &0.0036      \\ 
SE&0.0191      &0.0167      &0.1004      &0.0162      &0.0664      &0.0984      \\ 
95\% CI (normal)&(0.11,0.19) &(0.04,0.11) &(0.28,0.68) &(0.05,0.11) &(0.35,0.61) &(-0.19,0.2) \\ 
95\% CI (quantile)&(0.11,0.18) &(0.05,0.11) &(0.25,0.65) &(0.05,0.11) &(0.35,0.62) &(-0.23,0.17)\\ 
95\% CI (Fieller)&--          &--          &(0.27,0.68) &--          &(0.36,0.62) &--          \\ 
\hline
\multicolumn{1}{|l|}{}&\multicolumn{6}{c|}{Difference in fasting plasma glucose from baseline to $t_0$}\\ \hline
\multicolumn{1}{|l|}{}&\multicolumn{1}{c|}{$\Delta(t)$}&\multicolumn{1}{c|}{$\Delta_S(t,t_0)$}&\multicolumn{1}{c|}{$R_S(t,t_0)$}&\multicolumn{1}{c|}{$\Delta_T(t,t_0)$}&\multicolumn{1}{c|}{$R_T(t,t_0)$}&\multicolumn{1}{c|}{$IV_S(t,t_0)$}\\ \hline
Estimate1&0.1483      &0.0464      &0.6873      &0.0774      &0.4779      &0.2094      \\ 
SE1&0.0191      &0.0181      &0.098       &0.0162      &0.0664      &0.0585      \\ 
95\% CI (normal)1&(0.11,0.19) &(0.01,0.08) &(0.5,0.88)  &(0.05,0.11) &(0.35,0.61) &(0.09,0.32) \\ 
95\% CI (quantile)1&(0.11,0.18) &(0.01,0.08) &(0.52,0.9)  &(0.05,0.11) &(0.35,0.62) &(0.11,0.35) \\ 
95\% CI (Fieller)1&--          &--          &(0.52,0.91) &--          &(0.36,0.62) &--          \\ 
\hline
\end{tabular}
\vspace{3mm}
\end{center}
\caption{Proposed estimates examining two potential surrogate markers in the Diabetes Prevention Program:  the estimated treatment effect, $\widehat{\Delta}(t)$, the estimated residual treatment effect using surrogate information at $t_0$,  $\widehat{\Delta}_S(t,t_0) $, the estimated proportion of treatment effect explained by surrogate information at $t_0$, $\widehat{R}_S(t,t_0) $,the estimated residual treatment effect using survival information at $t_0$ only,  $\widehat{\Delta}_T(t,t_0) $, the estimated proportion of treatment effect explained by survival information only at $t_0$, $\widehat{R}_T(t,t_0) $, the incremental value of the surrogate marker information, $IV_S(t,t_0)$, with standard error (SE) estimates obtained using the perturbation-resampling procedure and 95\% confidence intervals (CI)  based on a normal approximation, a quantile-based calculation, and Fieller's  method where $t_0 = 1$ year and $t=3$ years \label{table-example}}
\end{table}

\end{document}